\documentclass[a4paper,11pt]{article}
\usepackage{pos}
\usepackage{graphicx}
\usepackage{bm}

\title{Inclusive processes from lattice QCD: problems and opportunities}

\author*[a,b]{Shoji Hashimoto}

\affiliation[a]{Institute for Particle and Nuclear Studies,
  High Energy Accelerator Research Organization (KEK),
  Tsukuba, Ibaraki 305-0801, Japan}

\affiliation[b]{SOKENDAI (The Graduate University for Advanced Studies),
  Tsukuba, Ibaraki 305-0801, Japan}

\emailAdd{shoji.hashimoto@kek.jp}

\abstract{
  A strategy to compute inclusive hadronic processes in lattice QCD is discussed. The key idea is to view the inclusive decay or scattering rate as a smeared spectrum. The Euclidean time dependence of correlators obtained on the lattice can be used to approximate them. The method induces its own systematic errors in addition to the standard discretization effects and so on. It is crucial to estimate them in a rigorous manner to achieve first-principles calculations of this important new class of quantities.
}

\FullConference{The XVIth Quark Confinement and the Hadron Spectrum Conference (QCHSC24)\\
 19-24 August, 2024\\
 Cairns Convention Centre, Cairns, Queensland, Australia\\}

\begin{document}
\maketitle

\section{Smeared spectrum and inclusive processes}
In the analysis of hadronic processes, there is a class of quantities that are written as an integral of spectral functions:
\begin{equation}
  \Gamma=\int_0^\infty\! d\omega \, K(\omega)\rho(\omega)
  \;\;\; \text{with} \;\;\;
  \rho(\omega)\propto \sum_X \delta(\omega-E_X) |\langle X|J|0\rangle|^2,
  \label{eq:smeared_spectrum}
\end{equation}
where the spectral function $\rho(\omega)$ characterizes the density of states having a given energy $\omega$, created by a current $J$ from the vacuum. The convolution kernel $K(\omega)$ specifies the weight of each state entering in the definition of $\Gamma$. (It can be a decay width or some other quantities.) Well-known examples include the hadronic vacuum polarization contribution to muon $g-2$ or the hadronic $\tau$-lepton decay rate, both of which can be defined with a sum of contributions from all possible final states $|X\rangle$ that can be generated from the vacuum $|0\rangle$. More generally, one can consider the case where the initial state is not the vacuum but a certain hadronic state. The inclusive semi-leptonic decay rate and inclusive $\ell N$ (lepton-nucleon) scattering cross section are examples of such quantities. 

Once the spectral function is obtained in the lattice QCD calculations, it can be used to reconstruct the integral \cite{Hansen:2017mnd}, but the computation of the spectral function is known as a notoriously difficult or ill-posed problem. Instead, one can consider a spectral function smeared over a range of $\omega$ to regularize the problem. Namely, the lattice calculation is performed for a series of finite smearing and the limit of unsmeared spectrum is approached \cite{Hansen:2019idp,Bailas:2020qmv}. One can go one step further by identifying the inclusive rate itself as a smeared spectrum and obtain the rate (or quantities built upon a certain integral of the differential rate) without recourse to the spectral function \cite{Hashimoto:2017wqo,Gambino:2020crt}.

In the lattice calculation, one typically computes correlation functions of some operators, {\it e.g.} a two-point correlator $C(t)=\langle 0|\tilde{J}(t)\tilde{J}(0)|0\rangle$ with some currents $J$ placed on the lattice with an Euclidean time separation $t$. (Fourier transform of the current $J$ is assumed in the spatial directions.) The correlator can be written as
\begin{equation}
  C(t) = \int_0^\infty\! d\omega\, \rho(\omega) e^{-\omega t}
  \label{eq:lattice_correlator}
\end{equation}
in the spectral representation. This can be viewed as an example of the smeared spectrum of (\ref{eq:smeared_spectrum}) with a smearing function $K(\omega)$ being $e^{-\omega t}$. In the operator representation, the correlator can also be written as $\langle 0|\tilde{J} e^{-\hat{H}t} \tilde{J}|0\rangle$ with the Hamiltonian operator $\hat{H}$ upon which the time-evolution operator $e^{-\hat{H}t}$ is built.

The similarity between the smeared spectrum (\ref{eq:smeared_spectrum}) and lattice correlator (\ref{eq:lattice_correlator}) suggests that the former can be approximated using a set of the latter. Indeed, if we write (\ref{eq:smeared_spectrum}) in the operator form as $\langle 0|\tilde{J}K(\hat{H})\tilde{J}|0\rangle$, one finds that the approximation can be established if the smearing kernel $K(\hat{H})$ is expanded as a polynomial of $e^{-\hat{H}}$:
\begin{equation}
  \label{eq:expansion}
  K(\hat{H})\simeq k_0+k_1 e^{-\hat{H}}+k_2 e^{-2\hat{H}}+\cdots+k_N e^{-N\hat{H}}.
\end{equation}
Here, I take the lattice unit, {\it i.e.} $a=1$.
Sandwiching by $\langle 0|\tilde{J}$ and $\tilde{J}|0\rangle$, the right-hand side of (\ref{eq:expansion}) is written by a linear combination of the lattice correlators $\langle 0|\tilde{J}e^{-\hat{H}t}\tilde{J}|0\rangle$ at $t$ = 0, 1, ..., $N$. Such an expansion is possible, when the kernel $K(\omega)$ is a slowly varying function. The problem then is to determine the optimal coefficients $k_j$. One strategy to obtain such an approximation was proposed in \cite{Hansen:2019idp} based on the Backus-Gilbert method as discussed in \cite{Hansen:2017mnd}. In this report, I focus on the Chebyshev approximation \cite{Bailas:2020qmv} and its properties. The numerical results are confirmed to be equivalent between these methods \cite{Barone:2023tbl}.

\section{Properties of Chebyshev approximation}
A systematic approach for the approximation can be constructed by the kernel polynomial method; see \cite{Weisse:2006zz} for example. An expansion of a function $f(x)$ defined in a certain range, say $x\in [-1,+1]$, in terms of orthogonal polynomials $p_n(x)$ is written as $f(x)=\sum_{n=0}^\infty \alpha_n p_n(x)$ with $\alpha_n=\langle p_n|f\rangle/\langle p_n|p_n\rangle$. The orthogonal polynomials are given once the inner product $\langle f|g\rangle=\int_0^1\!dx w(x)f(x)g(x)$ is defined with a weight function $w(x)$. Taking $w(x)=1/\pi\sqrt{1-x^2}$, one obtains the Chebyshev polynomials, defined by $T_0(x)=1$, $T_1(x)=x$, and $T_{m+1}(x)=2xT_m(x)-T_{m-1}(x)$. 

This Chebyshev-polynomial approximation
\begin{equation}
  \label{eq:Chebyshev}
  f(x)\simeq\frac{c_0}{2}+\sum_{j=1}^N c_jT_j(x)
\end{equation}
provides a nearly optimal approximation of the function $f(x)$ in the sense of {\it minmax}, {\it i.e.} maximum deviation is minimum for a given $N$. The convergence of (\ref{eq:Chebyshev}) is the fastest among other orthogonal polynomials. 
The coefficients $c_j$ are easily obtained up to arbitrary $N$ by performing a numerical integral of $\langle f|T_j\rangle$. Each term of the Chebyshev polynomials satisfies $|T_j(x)|\le 1$, so that the absolute upper limit of the ignored higher order terms $j>N$ is given as $\sum_{j=N+1}^\infty|c_j|$. Namely, the upper bound of the truncation error is known. A typical resolution of the Chebyshev polynomials truncated at $N$ is about $1/N$; the energy resolution for the kernel approximation is also $O(1/N)$. Therefore, in order to achieve better approximation, one needs larger $N$, {\it i.e.} the correlator at large time separations.

\begin{figure}[tbp]
  \centering
  \includegraphics[width=7.5cm]{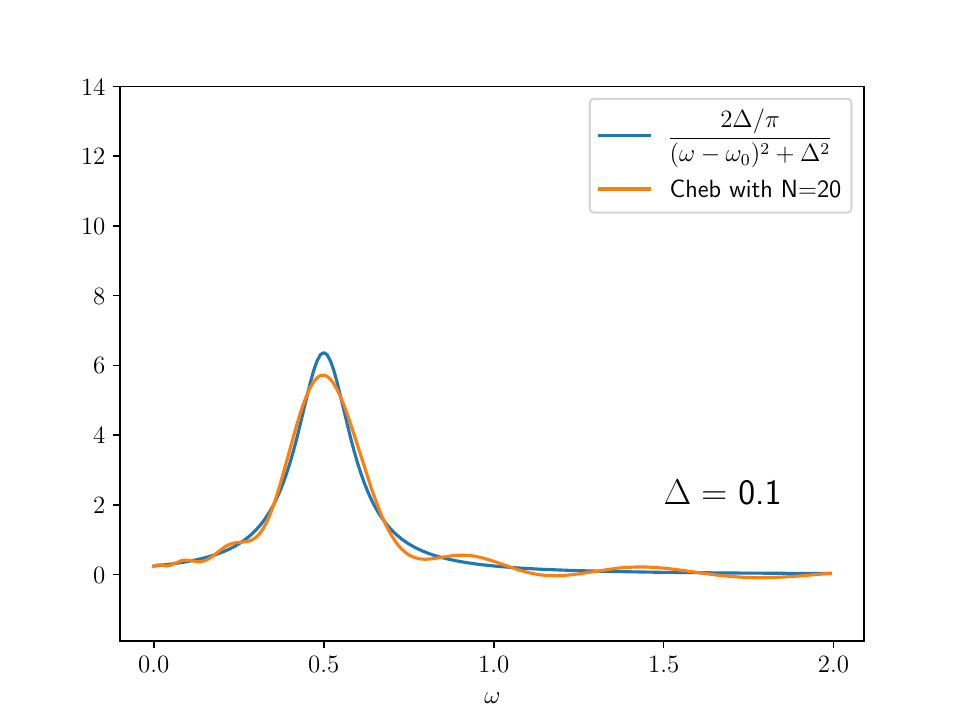}
  \includegraphics[width=7.5cm]{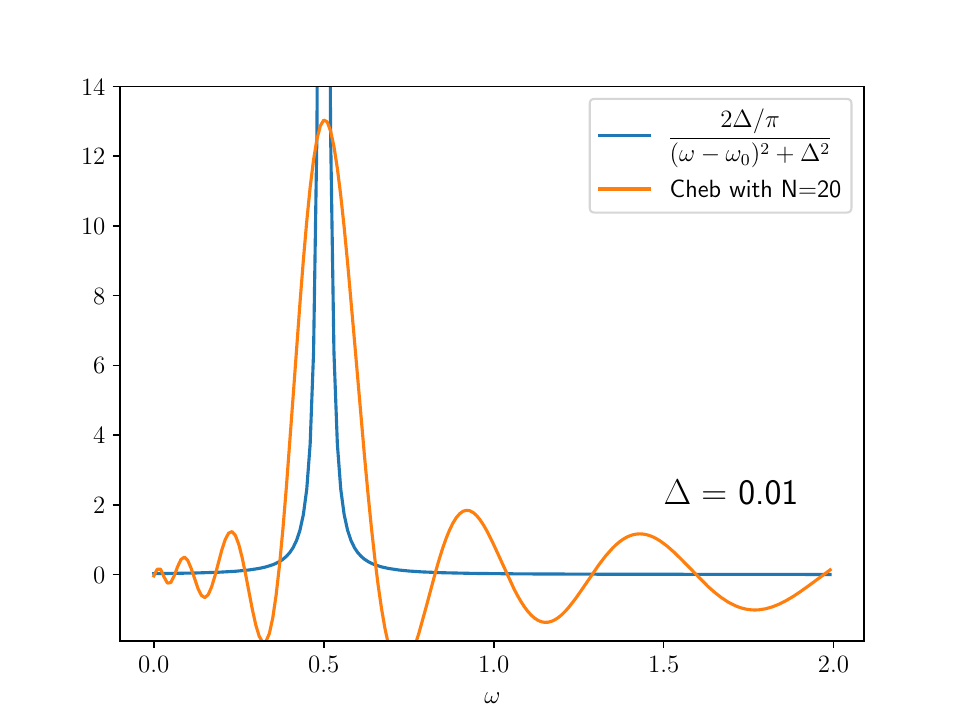}
  \caption{Approximation of a sharply peaked function $2(\Delta/\pi)/((\omega-\omega_0)^2+\Delta^2)$. The peak is at $\omega_0=0.5$, and the width $\Delta$ is 0.1 (left) and 0.01 (right). The Chebyshev approximation of the order $N=20$ is shown.}
  \label{fig:cheb_delta}
\end{figure}

A typical example of the Chebyshev-polynomial approximation is shown in Figure~\ref{fig:cheb_delta}. Here we take a smeared ``delta'' function $f(\omega)=2(\Delta/\pi)/((\omega-\omega_0)^2+\Delta^2)$ with the smearing width $\Delta$ = 0.1 (left) and 0.01 (right). The Chebyshev approximation at the order $N=20$ is shown for each of them. For $\Delta=0.1$, similar to the resolution $O(1/N)$, one can achieve fairly good approximation, while for $\Delta=0.01$ the approximation wildly oscillates and the deviation from the target function becomes substantial. To obtain a sensible approximation of the delta function, one therefore have to take the limit $\Delta\to 0$ together with (or after) $N\to\infty$. Or, the application of the approximation should be restricted to sufficiently smooth functions.

\begin{figure}[tbp]
  \centering
  \includegraphics[width=7.5cm]{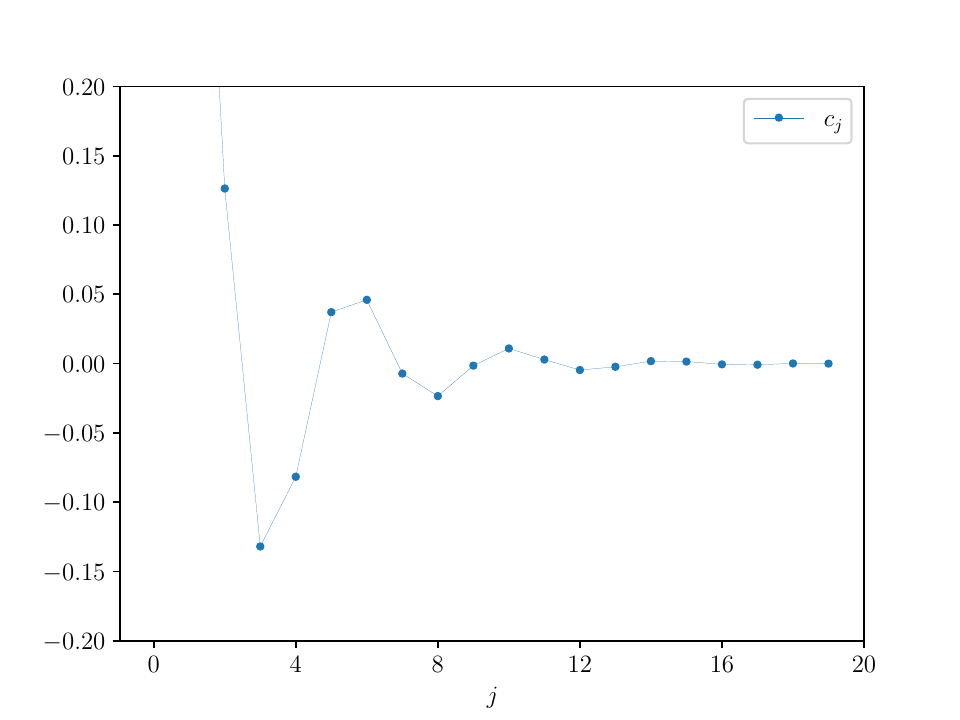}
  \includegraphics[width=7.5cm]{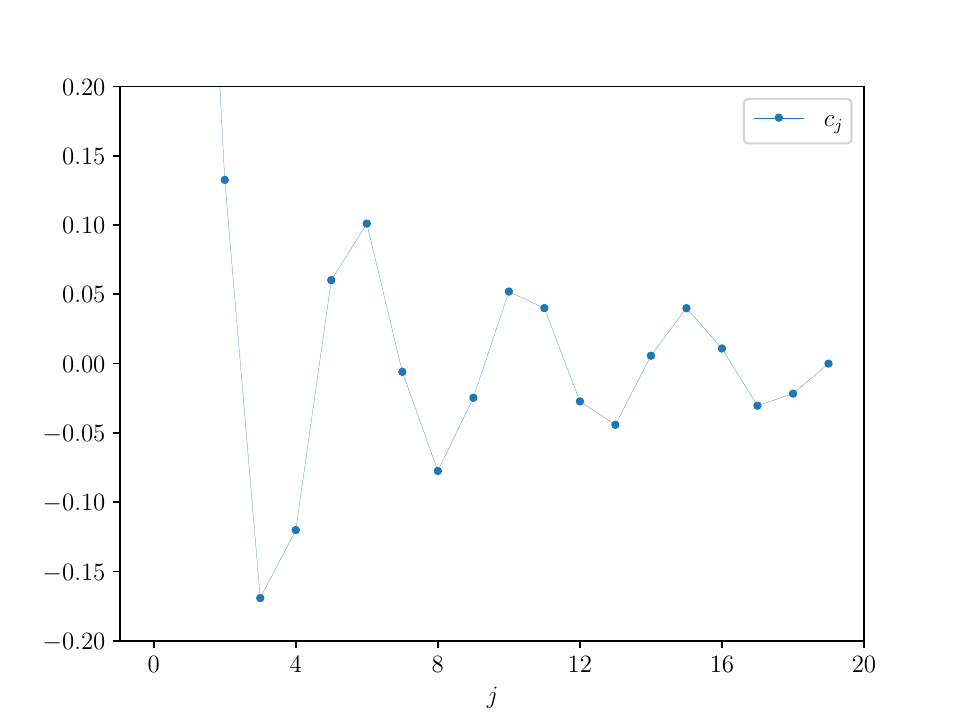}
  \caption{Chebyshev coefficients $c_j$ for the function $2(\Delta/\pi)/((\omega-\omega_0)^2+\Delta^2)$. The peak is at $\omega_0=0.5$, and the width $\Delta$ is 0.1 (left) and 0.01 (right).}
  \label{fig:cheb_coeff}
\end{figure}

The quality of the approximation can be monitored by its coefficients $c_j$'s. Figure~\ref{fig:cheb_coeff} shows the coefficients for $\Delta=0.1$ (left) and 0.01 (right). The convergence is clearly seen for $\Delta=0.1$ already around $j=12$, while the coefficient is still significantly oscillating even around $j=20$ for $\Delta=0.01$. In either case, the absolute value of $c_j$ decreases exponentially for larger $j$'s; its slope depends on $\Delta$ as $\sim e^{-\alpha\Delta j}$ with some numerical constant $\alpha$.

On the other hand, the Chebyshev matrix elements, $T_j(e^{-\hat{H}})$ sandwiched by desired states, are given in terms of the lattice correlators. For example, for the shifted Chebyshev polynomials\footnote{The shifted Chebyshev polynomials are defined in the range $x\in[0,1]$. It can also be adjusted for any finite range.} at $j=2$, $T_2^*(x)=8x^2-8x+1$, the matrix element $\langle T_2^*(e^{-\hat{H}})\rangle$ is given by $8\bar{C}(2)-8\bar{C}(1)+1$ from a normalized correlator $\bar{C}(t)=C(t+t_0)/C(t_0)$.

\begin{figure}[tbp]
  \centering
  \includegraphics[width=8cm]{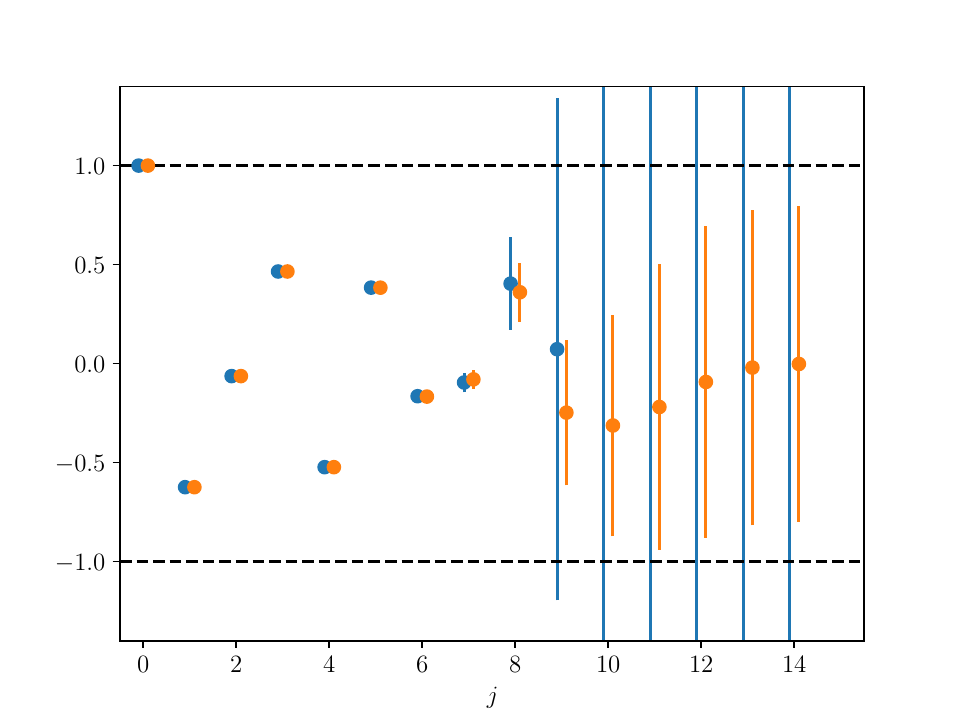}
  \caption{Chebyshev matrix elements $\langle T_j^*(e^{-\hat{H}})\rangle$ for the vacuum polarization function computed at pion mass 230~MeV. Naive reconstruction is shown by gray while the fit with the reverse formula with constraint $|\langle T_j^*(x)\rangle|$ is shown by orange.}
  \label{fig:Tj}
\end{figure}

An example is shown in Figure~\ref{fig:Tj}. The lattice data are from the 2+1-flavor domain-wall fermion simulation at $1/a$ = 2.45~GeV on a $48^3\times 96$ lattice. (The description of the ensemble is found in \cite{Colquhoun:2022atw}.) The sea/valence pion mass is $\sim$ 230~MeV, and the local-local vector current correlator is analyzed. The Chebyshev matrix elements can be constructed naively as described above, but the statistical error diverges for larger $j$'s beyond the constraint $|\langle T_j^*(x)\rangle|\leq 1$. Once the Chebyshev matrix element went out of the range $[-1,+1]$, the approximation breaks down, {\it i.e.} the result gets totally out of control. To avoid this problem, we introduce a constraint $|\langle T_j^*(x)\rangle|\leq 1$ when we determine the matrix elements $\langle T_j^*(e^{-\hat{H}})\rangle$. Namely, the matrix elements are obtained by a fit of the data using the reverse formula of Chebyshev polynomials together with the constraints. The details are discussed in \cite{Bailas:2020qmv}. With this prescription, the coefficients are confined within $[-1,+1]$; for large $j$'s, beyond $j\gtrsim 10$, the data do not have a constraining power and the matrix elements are simply given by the constraint. 

Let me share an interesting physics application of the Chebyshev approximation approach. It is a computation of the Borel transform of the spectral function, which is defined as
\begin{equation}
  \label{eq:Borel}
  \tilde{\Pi}(M^2)=\frac{1}{M^2}\int_0^\infty\! ds\,\rho(s)e^{-s/M^2},
\end{equation}
and often analyzed using the Operator Product Expansion (OPE) \cite{Shifman:1978bx,Shifman:1978by}. (Here the spectral density is defined as a function of $s=\omega^2$.) The Borel mass $M$ controls the range of the spectrum one focuses on. For sufficiently large $M$, the OPE is expected to converge well. On the lattice, we find that the Chebyshev polynomials approximation rapidly converges, and (\ref{eq:Borel}) is obtained essentially free of the truncation error.

\begin{figure}[tbp]
  \centering
  \includegraphics[width=10cm]{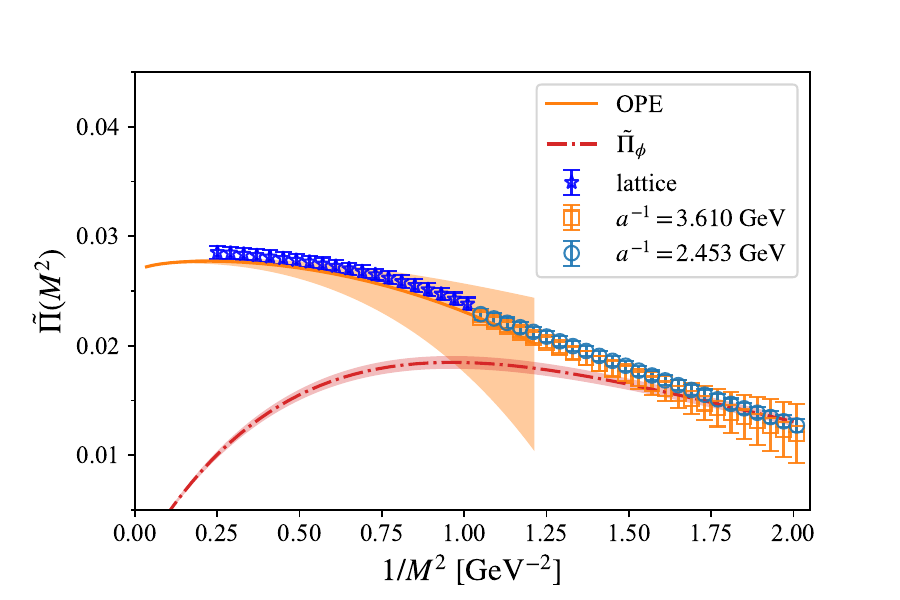}
  \caption{Borel transform of the vacuum polarization function in the $s\bar{s}$ channel plotted as a function of $1/M^2$. Lattice results are shown with symbols, while the estimate within OPE is given by a band. The dot-dashed curve represents the contribution of the ground-state $\phi$ meson. The plot is taken from \cite{Ishikawa:2021txe}.}
  \label{fig:Borel}
\end{figure}

A lattice result \cite{Ishikawa:2021txe} is plotted in Fig.~\ref{fig:Borel} as a function of $1/M^2$. We find a good agreement with the perturbative calculation of OPE in the high-energy region. On the other hand, in the low-energy region, say $1/M^2>$ 0.5~GeV$^2$, the uncertainty of OPE rapidly grows due to poorly known condensates. The lattice calculation can be extended toward lower energy region and finally it meets the curve of the ground-state ($\phi$ meson) contribution.

\section{Inclusive semileptonic decay as an example}
A more involved but concrete example of the calculation of the smeared spectral function has been given for the inclusive semileptonic decays of $B$ or $D$ mesons \cite{Gambino:2020crt,Gambino:2022dvu}. The inclusive decay rate of the $B$ meson is written as
\begin{equation}
  \label{eq:inclusive_rate}
  \Gamma\propto \int_0^{\bm{q}^2_{\text{max}}}\!d\bm{q}^2
  \int_{\sqrt{m_D^2+\bm{q}^2}}^{m_B-\sqrt{\bm{q}^2}}\!d\omega\,
  K(\omega;\bm{q}^2)
  \langle B(\bm{0})|\tilde{J}^\dagger(-\bm{q})\delta(\omega-\hat{H})\tilde{J}(\bm{q})|B(\bm{0})\rangle,
\end{equation}
where the integral is first performed for the energy $\omega$ of the hadronic final state and then for the recoil momentum $\bm{q}^2$. The kernel $K(\omega;\bm{q}^2)$ is given by the leptonic part of the amplitude and is explicitly known as a function of $\omega$ and $\bm{q}$. The lower and upper limit of the $\omega$ integral is also set by the kinematics of the semileptonic decay. The operators $\tilde{J}(\bm{q})$ represent the $b\to c$ flavor-changing currents, and the momentum $\bm{q}$ of the final hadronic states is inserted by a Fourier transform.

The $\omega$-integral of (\ref{eq:inclusive_rate}) can be formally performed to obtain $\langle B(\bm{0})|\tilde{J}^\dagger(-\bm{q})\bar{K}(\hat{H};\bm{q}^2)\tilde{J}(\bm{q})|B(\bm{0})\rangle$, where the modified kernel $\bar{K}(\omega;\bm{q}^2)$ has the form
\begin{equation}
  \label{eq:kernel}
  \bar{K}(\omega;\bm{q}^2) \sim
  e^{2\omega t_0}(m_B-\omega)^l \theta(m_B-\sqrt{\bm{q}^2}-\omega),
\end{equation}
taking account of the upper limit of the integral by the Heaviside function $\theta(m_B-\sqrt{\bm{q}^2}-\omega)$. The lower limit can be set to any value below $\sqrt{m_D^2+\bm{q}^2}$, since there is no hadronic state below this ground state energy. It is however advantageous to set the lower limit as large as possible to minimize the range of $\omega$ to be approximated \cite{Barone:2023tbl}.
The factor $(m_B-\omega)^l$ is from the above mentioned kinematical factor from the leptonic part of the amplitude, while the factor $e^{2\omega t_0}$ is introduced to keep a finite distance $t_0$ between the two inserted currents.

The kernel (\ref{eq:kernel}) involves a discontinuity introduced by the Heaviside function; the Chebyshev approximation of such function requires large $N$, the order of the polynomials, since the resolution of the approximation scales as $1/N$. Otherwise, the systematic error due to the truncation of the polynomials can become prohibitively large. We can regulate the problem by introducing a smearing of the Heaviside function and then taking the limit of vanishing width of the smearing.

\begin{figure}[tbp]
  \centering
  \includegraphics[width=7.5cm]{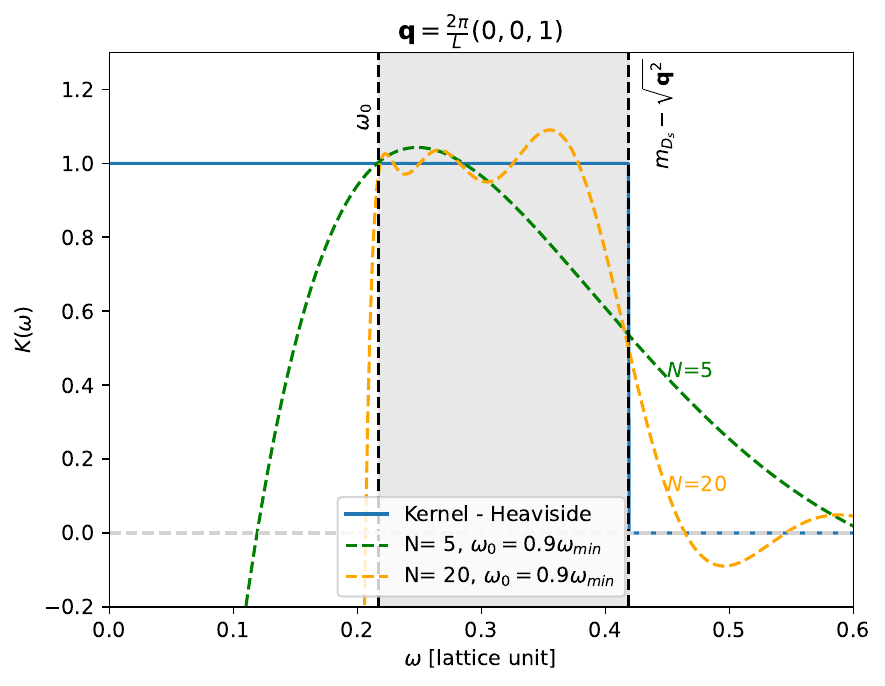}
  \includegraphics[width=7.5cm]{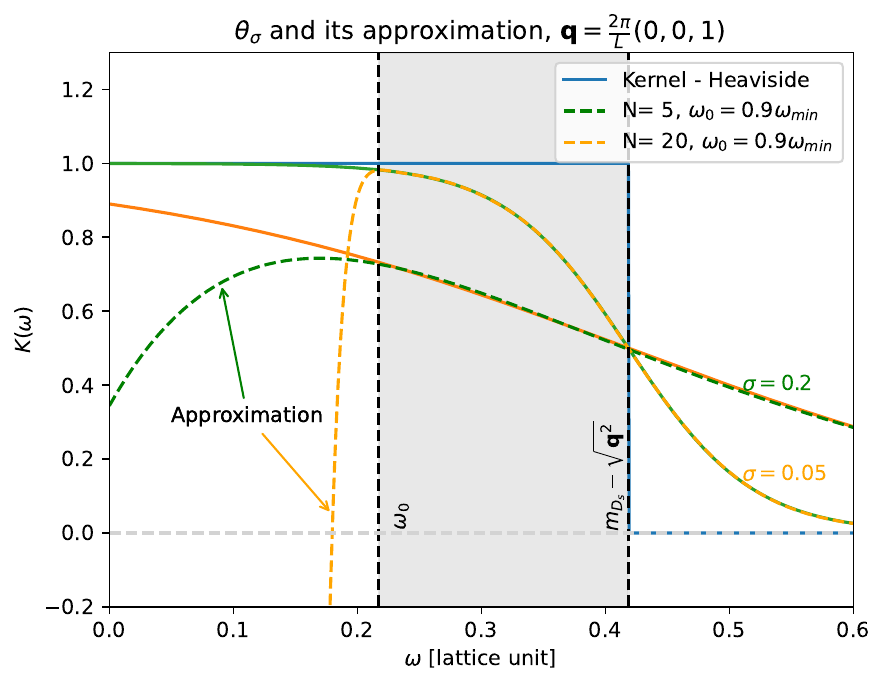}
  \caption{Chebyshev approximation of the Heaviside function (left) and its smeared version implemented by the sigmoid function (right). The polynomial order $N$ is 5 (green) and 20 (red). The plots are from \cite{Kellermann:2022mms}. }
  \label{fig:Heaviside}
\end{figure}

An example for the study of the $D_s$ meson decay \cite{Kellermann:2022mms} is shown in Figure~\ref{fig:Heaviside}. Here the approximations of the Heaviside function (left) and the sigmoid function (right) are shown. the extra factors such as $e^{2\omega t_0}$ to compose the kernel function are ignored for simplicity. The Chebyshev approximation is applied from the minimum energy $\omega_0$ to infinity with the order $N$ = 5 and 20. We find that the approximation follows the target function fairly precisely for the sigmoid function. For $\sigma=0.2$, the width of the smearing, the Chebyshev polynomial of order $N=5$ is sufficient to reproduce the function, while a larger $N$ is required for smaller $\sigma$, as anticipated from the scaling of the resolution $\sigma\sim 1/N$. The Chebyshev approximation of the same order shows significant oscillation around the target function when applied to the Heaviside function. For a controlled approximation, it would be necessary to keep $\sigma$ at the order of $1/N$ and to take the limit $\sigma\to 0$ while keeping $\sigma=1/N$.

Insufficient $N$ or finite $\sigma$ induces a systematic error in the evaluation of the inclusive decay rate, which is given by a convolution integral of the kernel function with the spectral function. The latter is determined by the dynamics of QCD, and we don't know its form a priori. The problem becomes severer for larger recoil momentum, because the allowed phase space is narrower.

\begin{figure}[tbp]
  \centering
  \includegraphics[width=7.5cm]{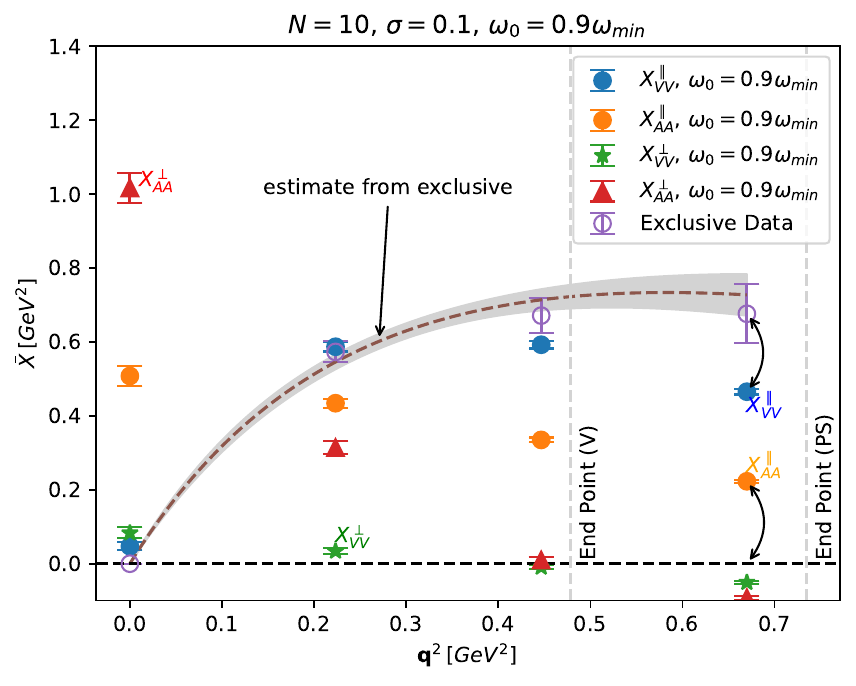}
  \includegraphics[width=7.5cm]{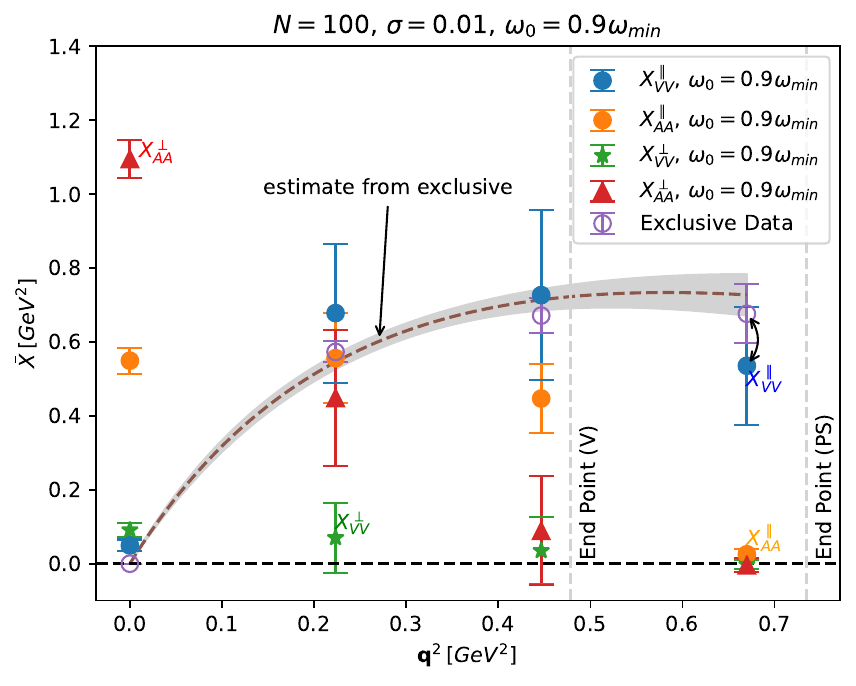}
  \caption{Differential decay rate (divided by $|\bm{q}|$ as a function of $\bm{q}^2$. Contributions of different channels are shown separately. The inclusive analysis is performed with $N=10$ (left) and 100 (right). The plots are from \cite{Kellermann:2022mms}.}
  \label{fig:diff_rate}
\end{figure}

The result of the inclusive analysis is plotted for the $D_s$ semileptonic decays in Figure~\ref{fig:diff_rate}. Contributions from different current insertions ($VV$ or $AA$; $\parallel$ or $\perp$) are shown separately as a function of momentum squared $\bm{q}^2$ of the final state hadrons. The vertical dashed lines indicate the kinematical endpoint depending on the lowest-energy final state particles (pseudo-scalar (PS) or vector (V)). The PS applies for the $VV\parallel$ channel while other channels are saturated by V. Also plotted by a gray band is an estimate of the ground-state contribution obtained from an independently measured $D_s\to\eta_s$ form factor, which is obtained in the course of \cite{Aoki:2023qpa}. 

We notice that the inclusive decay rate of the $VV\parallel$ channel is {\it lower} than the corresponding contribution from the ground state at the highest recoil momentum $\bm{q}^2\simeq$ 0.7~GeV$^2$. Also, the other channels should vanish beyond the kinematic endpoint (V) while the result for $AA\parallel$ is significantly higher than zero. These observations signal important systematic errors behind the method used in this analysis.

\section{Systematic errors}

\begin{figure}[tbp]
  \centering
  \includegraphics[width=8cm]{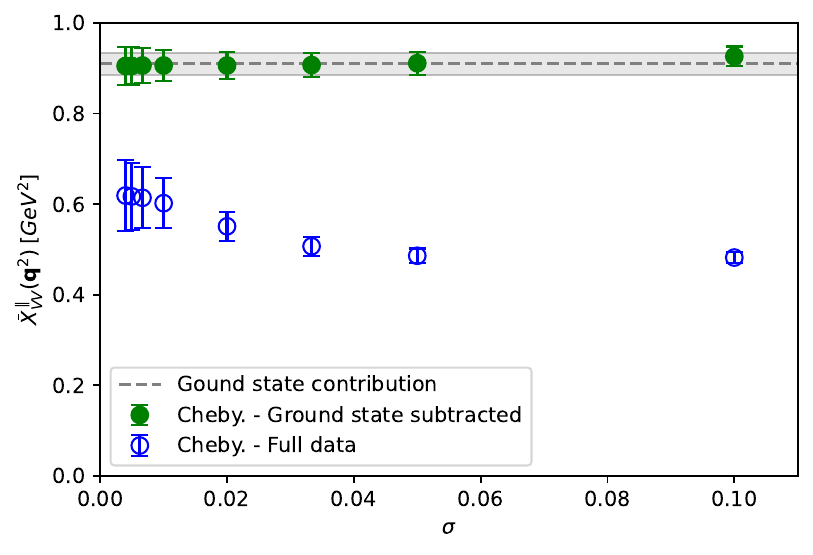}
  \caption{Extrapolation of $\bar{X}_{VV}^\parallel$ to the $\sigma\to 0$ limit. The smearing width is set to $\sigma=1/N$. Result of the inclusive analysis is shown by open symbols while the improvement by subtracting the ground-state contribution is represented by filled symbols. The data point is at the largest recoil momentum. The plot is taken from \cite{Kellermann:2024jqg}.}
  \label{fig:sigma_extrap}
\end{figure}

The problem of the poor approximation with the limited order of polynomials arises more significantly for large recoil momenta. Indeed, it turns out that the decay rate is largely underestimated when the energy of the lowest-lying state is just below the threshold where the Chebyshev approximation is poor.  Indeed, the estimate at a finite order $N$ depends strongly on $\sigma$ which is set to $\sigma=1/N$, as shown in Figure~\ref{fig:sigma_extrap} (blue circles). The error grows toward large $N$ because the corresponding Chebyshev matrix elements $\langle T_j^*(e^{-\hat{H}})\rangle$ is essentially unknown and distributes uniformly between $-1$ and $+1$. The error would further increase towards $\sigma\to 0$. This problem can be avoided by explicitly subtracting the low-lying state and treating it exactly. The subtraction can be done simply by a standard exponential fit of the correlator. Then, since the energy is fixed by the fit, the convolution integral with the kernel is trivial. The remaining contributions from higher energy states can then be included using the inclusive analysis. The result is shown in Figure~\ref{fig:sigma_extrap} (red circles). In contrast to the fully inclusive analysis, the result does not depend on $\sigma=1/N$.

Another important source of the systematic error for the inclusive decay rate would be the finite-volume effect, since the multi-body final states contribute and their finite-volume effect is expected to scale as an inverse power of the volume rather than a strongly suppressed exponential scaling. An explicit confirmation of such a power-like scaling is yet to be observed; only an estimate assuming some hadronic decay form factor is available. An analysis in \cite{Kellermann:2024jqg} suggests that the systematic error is not very significant for their setup where the final state kaons are heavier than their physical value.

\section{Summary}
To summarize, a systematic approach to compute the inclusive processes on the lattice is now available. The inclusive processes involve an integral over the energy of the hadronic state with a given weight (or a kernel), and it is approximated using the Euclidean time-dependent correlators. The method induces its own systematic effect such as the truncation of the kernel approximation, but the associated error can be estimated on a solid theoretical basis. Finite-volume effect is another important source of potential systematic error, which needs more studies.

Early analyses of the semileptonic decays include \cite{Gambino:2022dvu,Barone:2023tbl} for $D_{(s)}$ and $B_{(s)}$ meson decays, and more is expected in the near future, including more complete calculations that can be compared with the experimental data.

More applications can be considered for the smeared spectral function. The inclusive lepron-nucleon scattering can be treated in parallel to the semileptonic decays \cite{Fukaya:2020wpp}. Comparison with the experimental data for the $R$ ratio is found in \cite{ExtendedTwistedMassCollaborationETMC:2022sta}. Another prominent example is the hadronic $\tau$-lepton decays, for which extensive analyses have been performed recently \cite{Evangelista:2023fmt,ExtendedTwistedMass:2024myu}. These quantities can be constructed from lattice two-point correlators. Another class of interesting applications would be the extraction of decay/scattering amplitudes at various kinematics \cite{Bulava:2019kbi,Frezzotti:2023nun}, and more would be expected as the methods are developed in the future.

\vspace*{7mm}
I thank the present and past members of the JLQCD collaboration,
as well as the collaborators that led to the publications
\cite{Gambino:2022dvu,Barone:2023tbl}.
A lot of materials in this presentation emerged from the discussions
with them.

This work is partly supported by MEXT as ``Program for Promoting
Researches on the Supercomputer Fugaku'' (JPMXP1020200105)
and by JSPS KAKENHI, Grant-Number 22H00138.
This work used computational resources of supercomputer
Fugaku provided by the RIKEN Center for Computational Science
through the HPCI System Research Projects
(Project IDs: hp120281, hp230245),
SX-Aurora TSUBASA at the High Energy Accelerator Research
Organization (KEK) under its Particle, Nuclear and Astrophysics
Simulation Program
(Project IDs: 2019L003, 2020-006, 2021-007, 2022-006 and 2023-004).

\end{document}